\documentclass{article}
\usepackage{graphicx} 
\usepackage{longtable}
\usepackage{amsmath} 
\usepackage{booktabs} 
\usepackage{makecell}
\usepackage{subcaption}
\usepackage[table]{xcolor}

\title{Comprehensive Study of L-Menthol and Octanoic Acid as a Hydrophobic Eutectic Solvent}
\author{Bernarda Lovrin\v{c}evi\'{c}, Martina Po\v{z}ar, Romana Cerc Koro\v{s}ec,\\ Klara \v{S}vegelj, Peter Ogrin, Marija Be\v{s}ter-Roga\v{c}, Toma\v{z} Urbi\v{c}* }
\date{February 2026}

\begin{document}

\maketitle
\begin{abstract}
Hydrophobic eutectic solvents (HES) based on natural compounds represent promising green alternatives to conventional solvents. In this work, we investigate the physicochemical, structural, and dynamical properties of an ES formed by L-menthol and octanoic acid using a combined experimental and molecular dynamics simulation approach. Five compositions with molar ratios from 1:3 to 3:1 were studied with molecular dynamics simulation in the temperature range $15\:^\circ C-35\:^\circ C$. Experimental measurements of density and viscosity in the temperature range from $5\:^\circ C-35\:^\circ C$ were complemented with results obtained from MD simulations employing the OPLS force field. Structural analyses based on radial distribution functions and Kirkwood-Buff integrals reveals that the dominant interactions in the mixture are hydrogen bonds between L-menthol and octanoic acid molecules. Dynamic properties, including self-diffusion coefficients and hydrogen-bond lifetimes, indicate that intermolecular hydrogen bonds between the two components are stronger and longer-lived than bonds between identical species. 
These findings provide molecular-level insight into the structure and transport properties of menthol-based ESs relevant for green solvent applications.
\color{black}
\end{abstract}
\section{Introduction}
Eutectic solvents (ESs) have emerged as a promising class of green solvents, offering advantages such as low volatility, tunable properties, and environmental friendliness. They are typically formed by mixing a hydrogen bond acceptor (HBA) with a hydrogen bond donor (HBD), resulting in a eutectic mixture with a melting point significantly lower than that of the individual components. 
Among the various ESs, those based on natural compounds have garnered significant interest due to their biodegradability and low toxicity \cite{Smith2014}. \\L-menthol, a monoterpenoid alcohol derived from peppermint oil, serves as an effective HBD, while octanoic acid, a medium-chain fatty acid, acts as the HBA. The combination of L-menthol and octanoic acid forms a hydrophobic eutectic solvent (HES) with unique physicochemical properties making it suitable for industrial applications.

Phenolic compounds are known for their antioxidant, antimicrobial, anti-inflammatory, and other health-related effects \cite{Balasundram2006,Dai2010}. L-menthol, a naturally occurring monoterpene alcohol widely present in essential oils, also exhibits several biological activities, including antimicrobial and anti-inflammatory properties. However, phenolic compounds at high concentrations may be highly toxic for humans and the living environment; therefore, wastewater contaminated by phenolic compounds must be purified before discharge into the environment \cite{Khan2021}.
Specifically, eutectic mixtures formed by L-menthol with medium chain fatty acids can be successfully used in extraction processes \cite{Bergua2022, Cable2022} and produce better results than other types of ES. In addition, they can also be used to dissolve drugs that are otherwise unsolvable in water \cite{Bergua2022} and to extract important chemicals from plants \cite{Krizek2018, Wang2020} and other organisms \cite{Khare2021}.\\
Understanding the thermophysical properties of the L-menthol and octanoic acid ES is crucial for its effective application. 
Key properties include melting point, density, and viscosity, each influencing the solvent's behavior under various conditions. For instance, the density of such ESs has been reported to be lower than that of water \cite{Bergua2022}, with a temperature-dependent behavior that decreases linearly with increasing temperature.
Similarly, viscosity is a critical parameter affecting mass transfer and solubility, and it also exhibits temperature dependence \cite{Abbott2004,Bergua2022}, as well as surface tension which has been found to be very low, especially compared with similar ES and ionic liquids \cite{Nunes2019}.
Over the past few years, many computational studies have been performed on ES, mostly involving chloride-based hydrogen bond acceptors (HBA). Some authors focus on checking the force field parameters and improving them \cite{Kaur2019, Mainberger2017, Garcia2015, Salehi2019}, while others analyze the structure and dynamics driven by different interactions \cite{Hammond2017}, of which hydrogen bonding plays a key role. Gonzalez de Castilla et al. \cite{GonzalezdeCastilla2020} performed a study to test various models for ES coupled with MD simulations. Although they found the static and collective properties of ES to be represented fairly well, there are limits to describing the dynamic properties due to the polarizability issues \cite{Doherty2018, Mainberger2017}. However, classical force fields have been successfully used to compute and predict the vapor phase composition, solubility parameters and enthalpy of vaporization for choline-chloride based ES \cite{Salehi2019}.
A combined MD simulation and experimental study on choline chloride based eutectic mixtures by Perkins et al. demonstrated the ability of a classical force field to reproduce well experimental results, such as density, volume expansion coefficient and heat capacity \cite{Perkins2013,Perkins2014}. MD simulations are able to provide important information on industrially important processes in ES, since they can be used in biomass pretreatment \cite{Xu2021}. In addition, they can be easily employed to study particular biologically relevant processes, such as the stability and the function of lipase in ES \cite{Shehata2020}.
Coarse grained models, such as MARTINI, have been used for fatty-acid extraction and recovery process in imidazole-based ES \cite{Vainikka2025}, but are not suited for the description of the hydrogen bond dynamics. Phenyl-propionic acid and choline cloride ES have been studied via MD simulations with CHARMM force field \cite{JahanbakhshBonab2021}. Results have shown reasonable agreement with experimental values, as well as the structural description of the system in terms of the radial distribution functions and number of hydrogen bonds. ES composed of menthol and fatty acids has been studied in simulations with DFT-based force fields and the focus was on the effect of the acid chain length on structural and dynamic properties \cite{DabbaghHosseiniPour2025}. The same approach was taken for the analysis of water-in-ES system to explain how water disrupts the hydrogen bonding between the HBA and HBD \cite{Baranipour2023}. Malik and Kashyap \cite{Malik2021} have performed an extensive MD study on the structure of menthol-acid mixtures to investigate how the length of the acid alkyl tail affects the molecular structure, as seen from the structure factors $S(q)$. Their results indicate that polar correlations play a key role in the formation of the small-$q$ prepeak of $S(q).$  

In this study, we present a comprehensive analysis of the melting point, density, and viscosity of the L-menthol and octanoic acid ES at temperatures between 5\textdegree C and 55 \textdegree C, with an interval od 10 \textdegree C. Five types of ES have been studied with molar ratios of 1:3, 1:2, 1:1, 2:1 and 3:1. To elucidate the variations at the microscopic scale in these mixtures, induced by changes in temperature and concentration, we perform MD simulations on the aforementioned solvents for temperatures above the melting points. We analyze radial distribution functions (RDF), two-dimensional combined distribution functions (2D CDF), spatial distribution functions (SDF) and Kirkwood-Buff integrals (KBI) as structural properties and dynamical quantities such as self-diffusion coefficients ($D$). An additional analysis has been conducted involving the hydrogen bond dynamics. We calculated the average hydrogen bond lifetimes from the hydrogen bond auto-correlation function using a three-exponential fit. 
\color{blue}
This is, to the best of our knowledge, the first study to report Kirkwood–Buff integrals for this system, providing a quantitative measure of concentration fluctuations and preferential molecular associations within the mixtures. These results demonstrate preferential menthol–octanoic acid association, as seen by the least negative values of the cross KBIs, accompanied by suppression of self-association between like molecules. These findings provide direct molecular-level evidence for the intermolecular organization responsible for the stability of the eutectic mixture. Such structure–property relationships provide a quantitative basis for selecting compositions with the desired balance of intermolecular interactions, thereby reducing reliance on empirical trial-and-error approaches in the design of menthol-based eutectic solvents for applications such as extraction and formulation.
\color{black}
\section{Methods}
\subsection{Experimental setup}
\subsubsection{Materials}
L-menthol (C$_{10}$H$_{20}$O, MW = 156.27 g mol$^{-1}$, 99\%) and octanoic acid (C$_8$H$_{16}$O$_2$, MW = 144.21 g mol$^{-1}$, $\geq$ 98\%) were purchased from Sigma-Aldrich (Germany) and used as received. The studied systems were prepared by weighing L-menthol and octanoic acid on an analytical balance and stirring at moderate temperature ($\sim$ 20~$^\circ$C) until a homogeneous liquid was formed.

\subsubsection{Dynamic scanning calorimetry (DSC)}
\begin{figure}[h]
    \centering
    \includegraphics[width=0.9\textwidth]{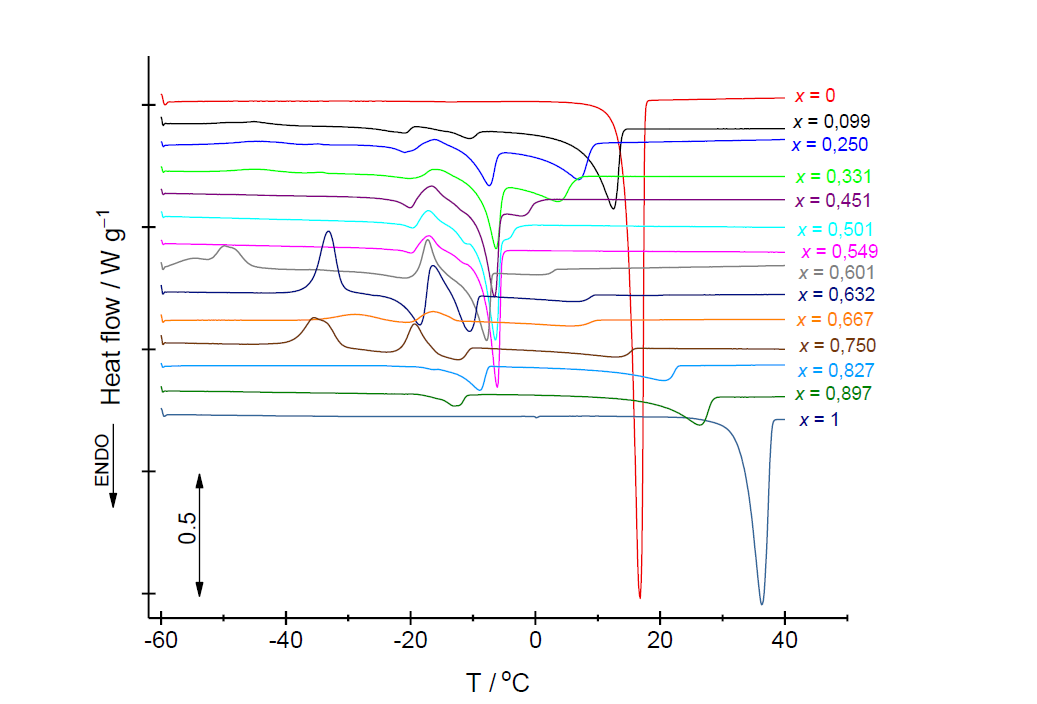} 
    \caption{DSC curves of pure octanoic acid ($x=0$), pure menthol ($x=1$), and their mixtures; $x$ represents the mole fraction of menthol.}
    \label{fig:dsc_curves}
\end{figure}

Differential scanning calorimetry (DSC) measurements were performed using a Mettler Toledo DSC instrument that had been calibrated beforehand. Temperature and enthalpy were calibrated with two standards: octane (reference substance for gas chromatography, Merck, Germany) and Milli-Q water. The corresponding liquid samples were placed in a 40~$\mu$L aluminium crucible using a micropipette, hermetically sealed and weighed on an external Mettler Toledo MX5 microbalance with an accuracy of 1~$\mu$g. The initial masses were between 5.0--6.9~mg. In the first ramp, the samples were cooled from room temperature to $-60\,^{\circ}\mathrm{C}$ at a cooling rate of $2~\mathrm{K\,min^{-1}}$, where they were thermally equilibrated for 10~minutes, and then heated to $60\,^{\circ}\mathrm{C}$ at a heating rate of $1~\mathrm{K\,min^{-1}}$. The resulting DSC curves for the investigated systems are shown in Figure \ref{fig:dsc_curves}.

\subsubsection{Density and viscosity measurements}
For density and viscosity measurements, five L-menthol/octanoic-acid mixtures were prepared with L-menthol to octanoic acid molar ratios of 1:3, 1:2, 1:1, 2:1, and 3:1.

Densities were measured over the temperature range 5~$^{\circ}$C to 55~$^{\circ}$C in 10~$^{\circ}$C increments using a vibrating-tube densimeter (Anton Paar DSA 5000 M, Austria). The instrument measures density with a repeatability of $\pm$0.000001~g\,cm$^{-3}$ and an error of $\pm$0.000007~g\,cm$^{-3}$, and temperature with a repeatability of $\pm$0.001~$^{\circ}$C and an error of $\pm$0.01~$^{\circ}$C.

The viscosity of the investigated ES was determined by measuring the flow rate of the sample in a Cannon--Fenske viscometer (SI Analytics GmbH, Mainz, Germany, type no.~520 20/200, $K = 0.09116$~mm$^{2}$\,s$^{-1}$). The flow time was measured using an attached automatic AVS~370 system. The temperature was controlled by immersing the viscometer in a polyethylene glycol bath equipped with a Lauda DLK~10 thermostat. Measurements were performed only after the temperature had been maintained at $\pm$0.01~$^{\circ}$C for 15~minutes. All flow-time measurements were repeated five times with a maximum permissible deviation of 1\%. 

The dynamic viscosity ($\eta$) was calculated using the equation:
\begin{equation}
\eta = \rho \cdot K \cdot t
\end{equation}
where $\rho$ is the density of the system, $t$ is the flow time, and $K$ is the viscometer constant ($K = 0.09116$~mm$^{2}$\,s$^{-1}$, as provided by the manufacturer and verified by calibration with standard D10 ISO~17025/ISO~17034).

\subsection{Simulation details}
\begin{figure}
    \centering
    \includegraphics[width=0.9\linewidth]{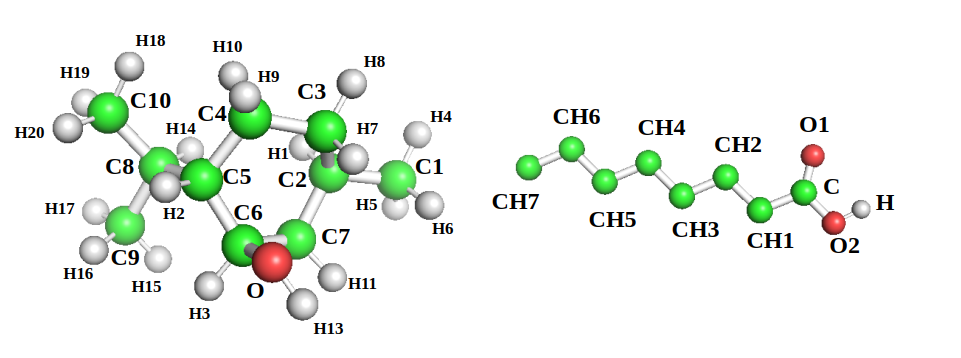}
    \caption{Schematic representation of L-menthol (left) and octanoic acid (right) with the corresponding site labels used in this study.}
    \label{fig:mols}
\end{figure}

All the simulations were performed using the GROMACS package version 2023.3 \cite{Pronk2013}. Each simulation box contained 2048 molecules in total. Initial configurations were generated using the Packmol program \cite{Martinez2003}. The simulation protocol was as follows. After energy minimization, the system first underwent equilibration in the NVT ensemble for 1 ns, then in the NPT ensemble for an additional 1 ns. This was followed by a 5 ns long production run in the NPT ensemble with the pressure set at $p=1\:atm$ and three different temperatures $T=288\: K,298\:K,308\:K$ to match experimental values. Parrinello-Rahman barostat \cite{Parrinello1980, Parrinello1981} with time constant of 5 ps was used to keep the pressure fixed. The thermostat was V-rescale \cite{Bussi2007} with time constant of 0.2 ps to stabilize the temperature. Leap-frog algorithm was used for the integration of the equations of motion with a time step of 2 fs and LINCS algorithm \cite{Hess1997} was used to keep the molecules stable. 
The cut-offs for all the interactions, including van der Waals and Coulomb, was set to 1.5 nm. The long-range electrostatics were calculated using the PME method \cite{Darden1993} with FFT grid size of 0.12 nm and spline interpolation of order 4. Molecular geometries for the two components with all the atomic sites are presented in Figure \ref{fig:mols}. 
\color{blue}For octanoic acid, we used the standard OPLS united-atom representation, in which the alkyl chain is described by the original OPLS-UA alkane parameters and the carboxylic acid head group by the parameters of Briggs, Nguyen, and Jorgensen \cite{Briggs1991} who validated this parameterization against experimental thermodynamic and structural properties of liquid acetic acid (agreement to within 1-4\%). This UA description is well established for saturated fatty acids and offers a substantial reduction in computational cost for the alkyl tail without sacrificing accuracy, since the aliphatic $CH_2/CH_3$ groups are not involved in the hydrogen-bonding interactions that govern the behavior of the mixture.
At the time the topologies for these compounds were assembled, there was no united-atom parameterization available in the literature that was specifically validated for menthol. We therefore use the OPLS-AA-compatible parameter set of Jasik and Szefczyk \cite{Jasik2016}, who developed and validated this force field specifically for menthol: partial charges, Lennard-Jones parameters, and torsional terms were refined against quantum-chemical reference data, and the resulting model was shown to reproduce the experimental density, enthalpy of vaporization, surface tension, and shear viscosity of pure liquid menthol over the temperature range $320-360\:K$. Regarding the concern that combining AA and UA representations could introduce inconsistencies in the inter-molecular interactions: both parameter sets are internally consistent members of the OPLS family and use the same non-bonded functional form and combining rules (i.e. geometric-mean $\sigma$ and $\epsilon$, 1-4 scaling of 0.5). Cross-interactions between menthol and octanoic-acid sites are therefore computed exactly as any other OPLS cross-term. Mixed-resolution setups of this kind are well-known within the OPLS framework, for example in membrane simulations, where united-atom lipid parameters are used alongside the all-atom OPLS-AA protein force field \cite{Ulmschneider2009}. This approach has been shown to give stable trajectories and physically sensible behavior for both components. Therefore, we don’t expect the AA/UA combination used here to introduce artifacts beyond those already inherent to any additive, fixed-charge force field.
The MD quantities are calculated as follows. We use the gmx rdf module to calculate the RDFs from the .xtc file with coordinates saved every 200 ps, as well as the diffusion coefficients via gmx msd module. The fitting of the mean-square displacement (MSD) curve is obtained from 300 ps to the last frame to ensure the diffusive regime. The spatial distribution functions were calculated via gmx spatial module with .gro and .xtc files as input.
Finally, the hydrogen bond autocorrelation function is obtained by using the gmx hbond module and the .trr file where the trajectories were saved every 200 ps.
\color{black}

\section{Results and discussion}
\subsection{Experimental results}
\subsubsection{Phase Behavior of L-menthol–Octanoic acid ES}

\begin{figure}[h]
\centering
\includegraphics[width=0.8\textwidth]{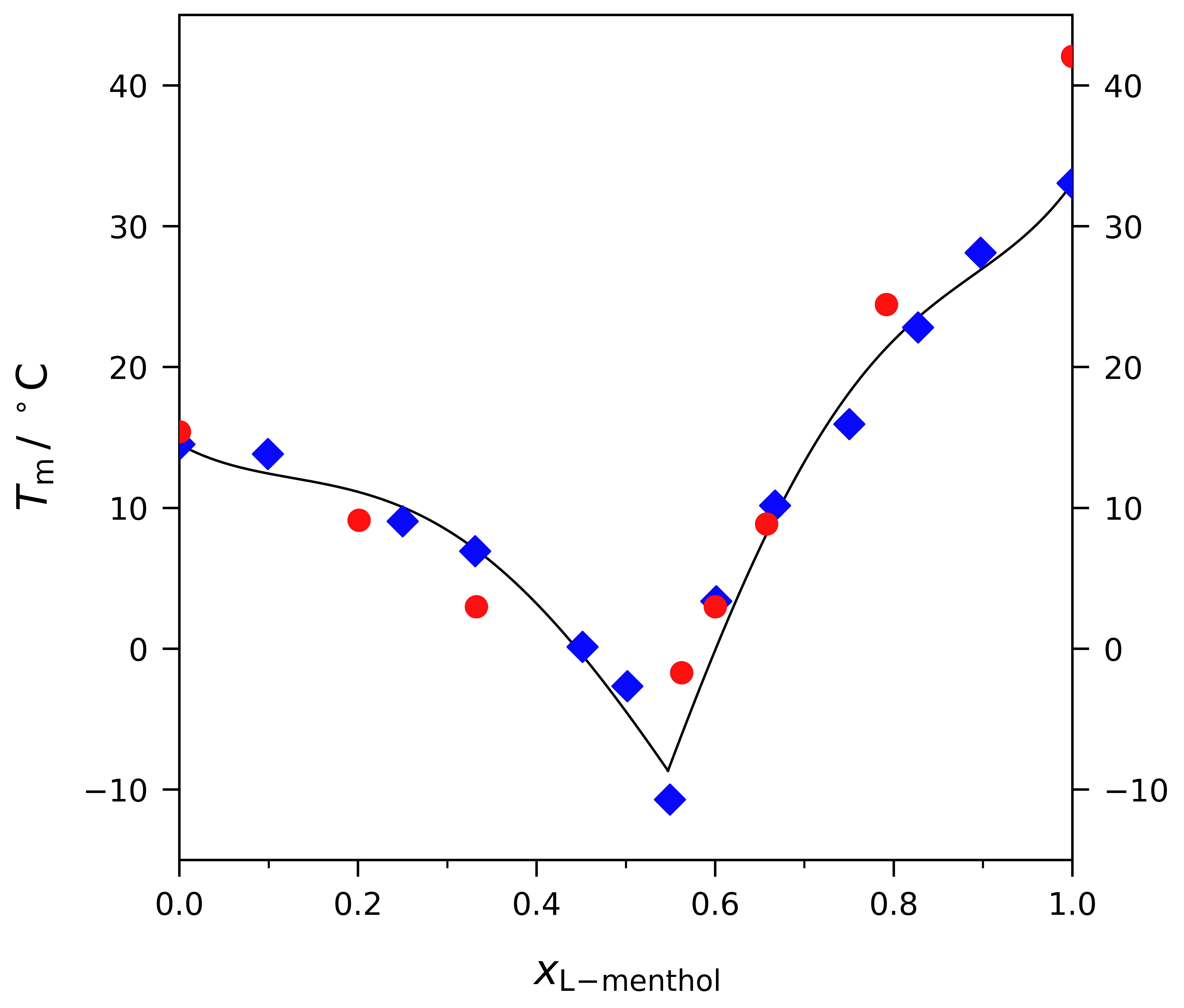}
\caption{Solid-liquid phase diagram of L-menthol–octanoic acid HES. Blue diamonds represent experimental data from this work; red circles represent data from literature \cite{Bergua2022}.}
\label{fig:phase_diagram}
\end{figure}

The melting point temperatures ($T_m$) for the L-menthol–octanoic acid system across various mol fractions of L-menthol ($x_L$) are summarized in Table \ref{tab:melting_points}. The resulting solid-liquid phase diagram (Figure \ref{fig:phase_diagram}) shows that the experimental data are in reasonable agreement with previously published results \cite{Bergua2022}.

In the present study, the eutectic point was identified at a molar fraction of $x_L = 0.549$ with a corresponding melting point of $T_m = -10.7$ $^\circ$C. This molar fraction is slightly higher than the previously reported value of $0.533$, while the observed melting point is lower than the literature value of $-9.53$ $^\circ$C \cite{Bergua2022}.

\begin{table}[h]
\centering
\caption{Melting point temperatures ($T_m$) for different mol fractions of L-menthol ($x_L$) in the investigated HES.}
\label{tab:melting_points}
\begin{tabular}{cc}
\hline
$x_L$ & $T_m$ ($^\circ$C) \\ \hline
0.0000 & 14.52 \\
0.0990 & 13.84 \\
0.2498 & 9.05 \\
0.3312 & 6.93 \\
0.4512 & 0.15 \\
0.5014 & -2.66 \\
0.5491 & -10.7 \\
0.6010 & 3.37 \\
0.6670 & 10.17 \\
0.7500 & 15.96 \\
0.8270 & 22.82 \\
0.8970 & 28.12 \\
1.0000 & 33.06 \\ \hline
\end{tabular}
\end{table}

\color{blue}
\subsubsection{Thermodynamic analysis of the solid-liquid equilibrium}
To quantify the non-ideal behavior of the L-menthol-octanoic acid system, the experimental liquidus data were analyzed using classical solid--liquid equilibrium thermodynamics. Assuming complete immiscibility of the solid phases and neglecting the heat-capacity difference between the solid and liquid phases, the liquidus temperatures satisfy the Schröder-van Laar equation:
\begin{equation}
\ln \left( x_i \gamma_i \right)
=
\frac{\Delta_{\mathrm{fus}} H_i}{R}
\left(
\frac{1}{T_{\mathrm{fus},i}}
-
\frac{1}{T}
\right),
\end{equation}
where $x_i$ and $\gamma_i$ denote the mole fraction and activity coefficient of component $i$, respectively, $T$ is the liquidus temperature, $T_{\mathrm{fus},i}$ is the melting temperature of the pure component, and $\Delta_{\mathrm{fus}}H_i$ is its molar enthalpy of fusion.
For an ideal liquid solution,
\begin{equation}
\gamma_i = 1,
\end{equation}
which gives the classical ideal liquidus.
To account for non-ideal behavior, the excess Gibbs energy of the liquid phase was described using the Redlich-Kister expansion:
\begin{equation}
\frac{G^{E}}{RT}
=
x_1 x_2
\left[
A_0
+
A_1 (x_1-x_2)
+
A_2 (x_1-x_2)^2
\right].
\end{equation}
The activity coefficients were calculated from
\begin{equation}
\ln \gamma_1
=
\frac{G^{E}}{RT}
+
x_2
\frac{\partial (G^{E}/RT)}{\partial x_1},
\end{equation}
\begin{equation}
\ln \gamma_2
=
\frac{G^{E}}{RT}
-
x_1
\frac{\partial (G^{E}/RT)}{\partial x_1}.
\end{equation}
The interaction parameters were determined by nonlinear least-squares optimization of the experimental liquidus temperatures.
The optimized model reproduces the experimental phase diagram well (Figure \ref{fig:phase_diagram}), yielding a root-mean-square deviation (RMSD) of approximately $2.5\:K$. The optimized interaction parameters indicate pronounced negative deviations from ideality throughout the composition range. Comparison of the optimized and ideal liquidus curves clearly demonstrates that the experimentally observed melting-point depression cannot be explained by ideal mixing alone.
The calculated activity coefficients at the eutectic composition are smaller than unity, confirming favorable heteromolecular interactions between L-menthol and octanoic acid. These attractive interactions stabilize the liquid phase and are responsible for the formation of the deep eutectic solvent.
\color{black}
\subsubsection{Density and Viscosity}
Experimental densities over the investigated temperature and composition ranges are presented in Table \ref{tab:densities}, where we compare the values with those previously reported in the literature \cite{Nunes2019}. The variation between the two sets of data is within the error bar and might be due to differences in sample preparation. 
Unsurprisingly, liquid densities decrease linearly with higher $T$ for all mole fractions $x_L$. If we compare densities with various $x_L$ while keeping the temperature $T$ fixed, we find that the values decrease linearly as one moves to the alcohol rich side of the mixture. Since the pure octanoic acid is slightly denser than the pure L-menthol all across our temperature range, this trend is expected. \color{blue}We show the values of MD densities and the absolute relative deviations for all three temperatures $T$ and five concentrations $x_L$ in the SI document. As shown in Table SI-1, these densities agree well with our experimental data. However, the calculated data show slightly overestimated values, with errors varying from $|\Delta \rho|=4.66\% - 8.54\%$. Our neat component densities are as follows. For octanoic acid: $\rho_{SIMU} = 0.9835\:gcm^{-3}$ at $T = 298\:K$ and $\rho_{SIMU} = 0.9785\:gcm^{-3}$ at $T = 308\:K$ (our experimental value is $\rho_{EXP} = 0.8982\:gcm^{-3}$ causing the error $|\Delta \rho| = 8.94\%$). 
For L-menthol: $\rho_{SIMU} = 0.9074\:gcm^{-3}$ at $T = 298\:K$ and $\rho_{SIMU} = 0.8997\:gcm^{-3}$ at $T = 308\:K$ (our experimental value is $\rho_{EXP} = 0.8866\:gcm^{-3}$ which causes the error $|\Delta \rho| =1.5\%$).
\color{black}


\begin{table}[h]
\centering
\caption{Density ($\rho$) for 7 different mole fractions of L-menthol between 0 and 1 over the temperature range from $5 ^\circ C$ to $55 ^\circ C$. Values in parenthesis are from Ref. \cite{Nunes2019}.}
\label{tab:densities}
\begin{tabular}{cccccccc}
\hline
\multicolumn{8}{c}{\cellcolor{gray!20}$\rho$/g$\cdot$cm$^{-3}$} \\
\hline
\cellcolor{gray!20}$T/^\circ C$ & \cellcolor{gray!20} $x_L=0$ & \cellcolor{gray!20}$x_L=0.2498$ & \cellcolor{gray!20}$x_L=0.3312$ &\cellcolor{gray!20} $x_L=0.5014$ & \cellcolor{gray!20}$x_L=0.6670$ & \cellcolor{gray!20}$x_L=0.7500$ & \cellcolor{gray!20}$x_L=1$ \\
\hline
5 & --     & --     & 0.9183 & 0.9162 & 0.9139 & 0.9127 & --    \\
\hline
15 & --     & 0.9116 & 0.9107 & 0.9087 & 0.9066 & 0.9053 & --    \\
\hline
25 & --     & \makecell{0.9038\\{\small(0.9147)}} & \makecell{0.9030\\{\small(0.9141)}} & \makecell{0.9013\\{\small(0.9123)}} & \makecell{0.8992\\{\small(0.9103)}} & \makecell{0.8980\\{\small(0.9093)}} & --    \\
\hline
35 & \makecell{0.8982\\{\small(0.9038)}} & \makecell{0.8961\\{\small(0.9017)}} & \makecell{0.8954\\{\small(0.9011)}} & \makecell{0.8937\\{\small(0.8995)}} & \makecell{0.8918\\{\small(0.8976)}} & \makecell{0.8906\\{\small(0.8966)}} & \makecell{0.8866\\{\small(0.8922)}} \\
\hline
45 & \makecell{0.8901 \\{\small(0.8917)}}& \makecell{0.8883\\{\small(0.8899)}} & \makecell{0.8877\\{\small(0.8894)}} & \makecell{0.8862\\{\small(0.8879)}} & \makecell{0.8843\\{\small(0.8861)}} & \makecell{0.8832 \\{\small(0.8850)}} & \makecell{0.8793\\{\small(0.8810)}} \\
\hline
55 & \makecell{0.8821 \\ {\small(0.8809)}}& \makecell{0.8806\\{\small(0.8795)}} & \makecell{0.8800\\{\small(0.8789)}}& \makecell{0.8786\\{\small(0.8776)}}& \makecell{0.8768\\{\small(0.8759)}} & \makecell{0.8757\\{\small(0.8744)}} & \makecell{0.8719 \\{\small(0.8708)}} \\
\hline
\end{tabular}
\end{table}

Next, we focus our attention on dynamic viscosity $\eta$ presented in Table \ref{tab:viscosities} for all temperatures $T$ and mole fractions $x_L$. We note that for all temperatures, neat L-menthol is more viscous than neat octanoic acid mostly due to its molecular shape. L-menthol is a bulky, rigid molecule with a cyclic hydrocarbon that creates steric hindrance and rotational friction when molecules try to flow past one another. Octanoic acid, on the other hand, has a relatively flexible linear chain which can improve the molecular flow, thus reducing the dynamic viscosity. L-menthol has a melting point of $T_m=33.06 ^\circ C$, which is close to room temperature, while octanoic acid melts much lower, at $T_m=14.52 ^\circ C$. This implies that L-menthol molecules have much less molecular motion near room temperature than acid ones, which is reflected in higher viscosity. Upon increasing L-menthol mole fraction $x_L$, the mixture becomes more viscous at all temperatures, as seen in Table \ref{tab:viscosities}. \color{blue} We did not calculate the MD viscosities from the present study since we expect that substantially longer simulation times would be required to obtain reliable estimates of this property. Instead, the present study focuses on the hydrogen-bond dynamics, diffusion, and structural properties of these systems, as characterized by MD simulations and complemented by experimental results.\color{black}

\begin{table}[h]
\centering
\caption{Dynamic viscosity ($\eta$) for 7 different mole fractions of L-menthol between 0 and 1 over the temperature range from $5\:^\circ C$ to $55\:^\circ C$. Values in the parenthesis are from Ref. \cite{Nunes2019}.}
\label{tab:viscosities}
\begin{tabular}{cccccccc}
\hline
\multicolumn{8}{c}{\cellcolor{gray!20}$\eta$/mPa$\cdot$s} \\
\hline
\cellcolor{gray!20}$T/^\circ C$ & \cellcolor{gray!20} $x_L=0$ & \cellcolor{gray!20}$x_L=0.2498$ & \cellcolor{gray!20}$x_L=0.3312$ &\cellcolor{gray!20} $x_L=0.5014$ & \cellcolor{gray!20}$x_L=0.6670$ & \cellcolor{gray!20}$x_L=0.7500$ & \cellcolor{gray!20}$x_L=1$ \\
\hline
5 & --     & --     & 20.845 & 34.75 & 67.11 & 104.0 & --    \\
\hline
15 & -- & 11.040 & 13.124 & 19.699 & 32.50 & 45.6 & -- \\
\hline
25 & -- & \makecell{7.700\\{\small(7.61)}} & \makecell{8.845\\{\small(8.81)}} & \makecell{12.200\\{\small(12.18)}} & \makecell{17.93\\{\small(17.71)}} & \makecell{22.3 \\{\small(20.82)}} & -- \\
\hline
35 & \makecell{4.11 \\ {\small(4.01)}}& \makecell{5.579\\{\small(5.54)}} & \makecell{6.248\\{\small(6.23)}}& \makecell{8.054\\{\small(8.06)}}& \makecell{10.79\\{\small(10.73)}} & \makecell{12.62\\{\small(11.94)}} & \makecell{24.4 \\{\small(23.19)}} \\
\hline
45 & \makecell{3.30 \\ {\small(3.20)}}& \makecell{4.21\\{\small(4.18)}} & \makecell{4.593\\{\small(4.60)}}& \makecell{5.64\\{\small(5.63)}}& \makecell{7.04\\{\small(7.00)}} & \makecell{7.87\\{\small(7.53)}} & \makecell{12.5 \\{\small(12.00)}} \\
\hline
55 & \makecell{2.707 \\ {\small(2.609)}}& \makecell{3.294\\{\small(3.25)}} & \makecell{3.52\\{\small(3.51)}}& \makecell{4.12\\{\small(4.11)}}& \makecell{4.87\\{\small(4.83)}} & \makecell{5.26\\{\small(5.05)}} & \makecell{7.20 \\{\small(6.92)}} \\
\hline
\end{tabular}
\end{table}

\subsection{Simulation results}
\subsubsection{Structural quantities}
\begin{figure}
    \centering
    \includegraphics[width=0.9\linewidth]{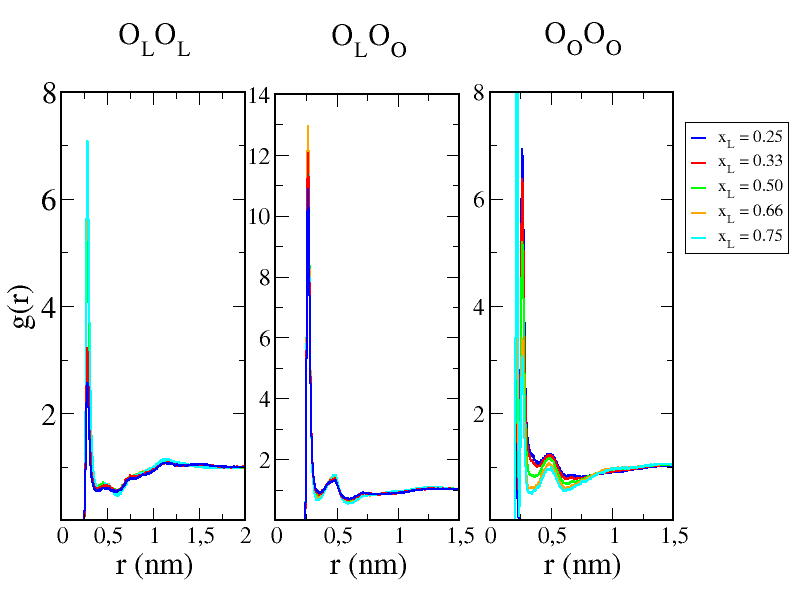}
    \caption{Radial distribution functions $g(r)$ between OO atoms at room temperature. Left panel: O-O (L-menthol), Middle panel: O-O2 (L-menthol-octanoic acid), Right panel: O1-O2 (octanoic acid). Color convention: $x_L=0.75$ (cyan), $x_L=0.66$ (orange),$x_L=0.50$ (green),$x_L=0.33$ (red), and $x_L=0.25$ (blue). }
    \label{fig:rdfOO}
\end{figure}

First, we discuss the structural properties of these systems in terms of the RDFs and the CNs. 
As seen in the left panel of Figure \ref{fig:rdfOO}, a very narrow oxygen-oxygen peak around $r=0.3\: nm$ between two L-menthol molecules becomes more prominent as we approach the alcohol-rich side of the mixture. This peak is typically found in mixtures where two oxygens are hydrogen bonded, such as in water-alcohol \cite{Mijakovic2014}. 
There is a depletion after the first peak that can be attributed to the chain-like formation of L-menthol molecules. The OO correlations between L-menthol and octanoic acid are depicted in the middle panel of Figure \ref{fig:rdfOO}. The first peak is again found at $r=0.3\: nm$, but there is very little variation in height across different concentrations. 
\color{green}This peak has the largest height, which leads to the conclusion that the main structural mechanism in the system is driven by the interaction between the two species. \color{black} 
On the right panel, we observe the OO correlations between two molecules of octanoic acid. As the acid becomes the minority component in the mixture, the correlations between like molecules become less prominent. This is in clear contrast with the $g(r)$ in the middle panel. There is a small local maximum at $r\approx0.5\: nm$ suggesting that a second neighbour shell forms due to the shape of the molecule, after which there is a depletion in the $g(r)$. A strong peak at $r\approx0.25\: nm$ is due to the intermolecular bond between the O1 and O2 atoms in the octanoic acid. 
In Figures SI-1. and SI-2. of the Supporting Information document, we show the same correlations for temperatures $T=15\:^\circ C$ and $35\:^\circ C$.

\begin{figure}
    \centering
    \includegraphics[width=0.9\linewidth]{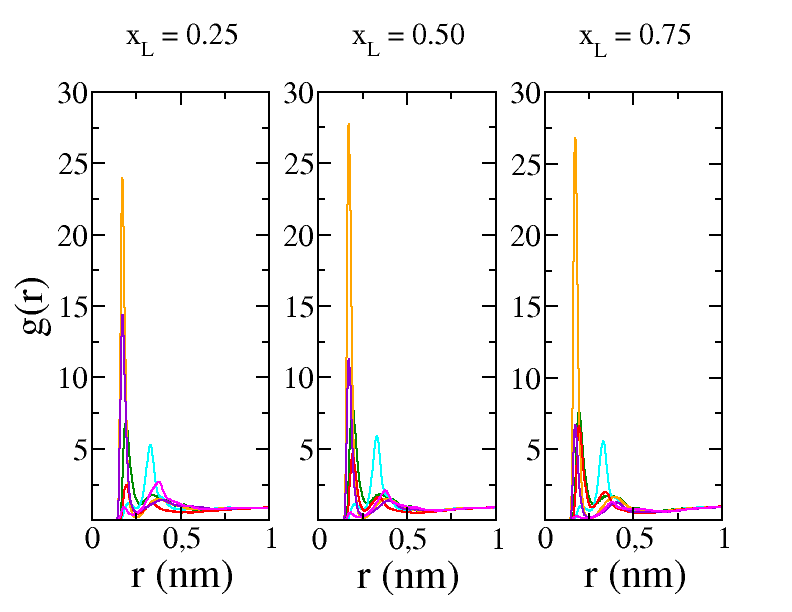}
    \caption{Radial distribution functions $g(r)$ between oxygen and hydrogen atoms at room temperature. Left panel: $x_L=0.25$, middle panel: $x_L=0.50$, right panel: $x_L=0.75$. Color convention: O (L-menthol)-H (octanoic acid) (orange), O1 (octanoic acid)-H (L-menthol) (green), O2 (octanoic acid)-H (L-menthol) (cyan), O (L-menthol)-H (L-menthol) (red), O1 (octanoic acid)-H (octanoic acid) (violet), O2 (octanoic acid)-H (octanoic acid) (magenta). }
    \label{fig:rdfOH}
\end{figure}

\color{blue}
In order to characterize hydrogen bonding between the two species, it is instructive to monitor the RDF of the oxygen and the hydrogen atoms. All possible OH RDFs are shown in Figure \ref{fig:rdfOH} at room temperature and for three systems with different amounts of L-menthol. As seen in all panels, the OH between the L-menthol O and the octanoic acid H has the strongest first neighbour correlation indicating the dominance between this hydrogen bonding pair. This peak is followed by the acid-acid RDF between the O1 atom and the H atom (violet curve) and the slightly shifted cross RDF between the O1 atom and the H atom on L-menthol. The L-menthol-L-menthol RDF between O and H has the smallest first peak, which is consistent with the fact that the O atom preferably bonds with the H atom from octanoic acid. The two remaining RDFs, the cross O2-H (magenta curve) and acid-acid O2-H (cyan) have the second neighbour peak more dominant than the first neighbour peak. This happens because of steric effects as the two sites cannot come closer due to the hydrogen bond formation between the cross O-H and the acid-acid O1-H pairs. The cross O-H correlations are almost unchanged with the addition of more L-menthol as the acid molecules preferably bond to alcohol regardless of the concentration.
The strongest difference is observed for the acid-acid O1-H RDF (violet curve) and L-menthol-L-menthol O-H RDF (red curve). As the acid becomes the minority component, correlations between molecules become weaker and those between alcohol molecules grow stronger. RDFs between O and H atoms for $T=15\:^\circ C$ and $T=35\:^\circ C$ presented in Fig. SI-3. and SI-4. in the Supporting Information document show very similar trends as the ones at room temperature.

\color{black}

\begin{figure}
    \centering
    \includegraphics[width=0.9\linewidth]{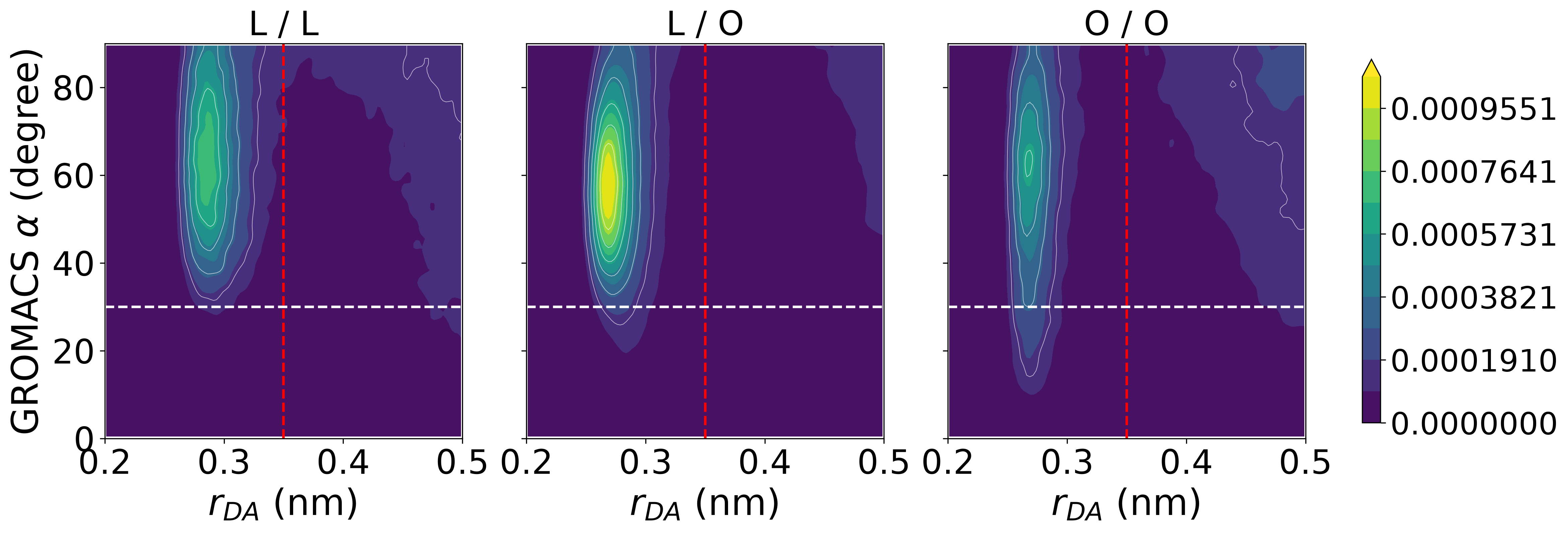}
    \caption{Two-dimensional combined distribution functions (2D CDFs) for $x_M=0.50$ at room temperature. These are based both on donor-acceptor distance $r_{DA}$ and the (H--D--A) angle $\alpha$ between L-menthol molecules (left panel), between L-menthol and octanoic acid (middle panel) and between octanoic acid molecules (right panel).  Color scale on the right corresponds to the normalized probability values. Dashed red and white lines indicate the limit distance $r_{DA} = 0.35\,\mathrm{nm}$ and the limit angle $\alpha=30^\circ$, respectively.}
    \label{fig:2dcdf}
\end{figure}

\color{green}
 Note that the RDFs provide information only regarding the observed distance between the two O-O and O-H sites, so they alone cannot be used as evidence for the hydrogen bonding inside the system. Therefore, we analyze the 2-dimensional combined distribution functions (2D CDFs) based on both the hydrogen-bond distance and angle criteria. We use the geometric definition for the hydrogen bond: a donor--acceptor (D--A) distance $r \leq 0.35\,\mathrm{nm}$ and a hydrogen--donor--acceptor (H--D--A) angle $\alpha \leq 30^\circ$ \cite{Luzar1996Nature, Luzar1996PRL}. Hydrogen-bond geometries were analyzed using the MDAnalysis Python package \cite{MichaudAgrawal2011, Gowers2016}. Our code is set up to collect all D--A contacts from 0 to 0.5 nm and all the angles up to $90^\circ$. 2D CDFs for $x_M=0.50$ at room temperature are presented in Figure \ref{fig:2dcdf}. All three distributions have a very strong maximum around distance $r\approx 0.27-0.26\:nm$, which is consistent with the first peak position of the $g(r)$ in Fig. \ref{fig:rdfOO}, but at the angle $\alpha\approx 55-65^\circ$. This suggests that the vast majority of our O-O contacts do not satisfy the conventional hydrogen bond angle criterion. O-O contacts of two octanoic acid molecules have the highest relative population of hydrogen bonds, as the rectangle bonded by the dashed lines at $r_{DA} = 0.35\,\mathrm{nm}$ and $\alpha=30^\circ$ contains the largest distribution area (right panel of Fig \ref{fig:2dcdf}). Clearly, octanoic acid molecules have a strong tendency to form directional associations.
 L-menthol and octanoic acid have the highest number of contacts and the highest overall number of hydrogen bonds, but the number of hydrogen bonds per contact is less than for the pair of octanoic acid molecules, as seen from the middle panel. Finally, L-menthol molecules are the weakest in terms of hydrogen bonding (left panel), thus confirming that the cross interactions drive the structural organization inside the mixture. Number of donor-acceptor contacts and hydrogen bonds for all three pairs are presented in Table SI-2. of the Supporting Information document.
\color{black}

\color{black}

\begin{figure}
    \centering
    \includegraphics[width=0.9\linewidth]{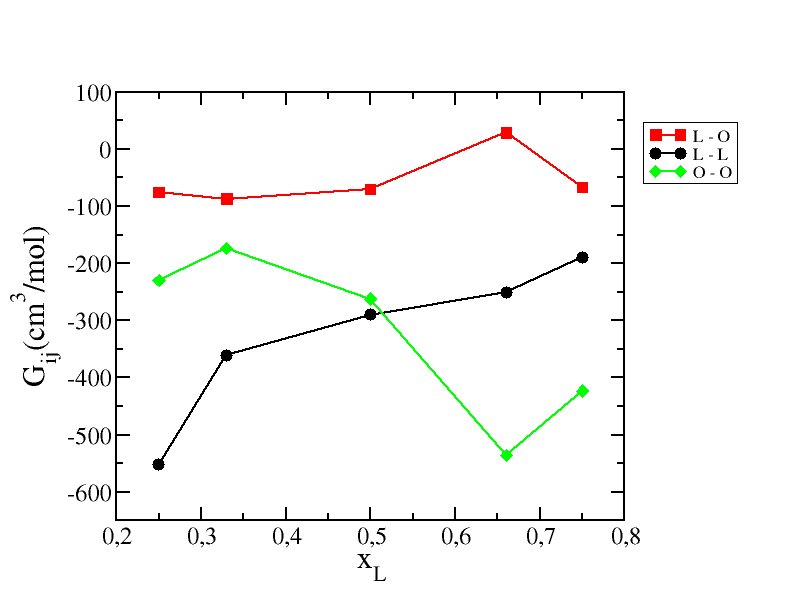}
    \caption{Kirkwood-Buff integrals $G_{ij}(cm^3/mol)$ vs L-menthol mole fraction $x_L$ at room temperature. Color convention: L-menthol-L-menthol (black circles), L-menthol-octanoic acid (red squares), octanoic acid-octanoic acid (green diamonds). }
    \label{fig:kbi}
\end{figure}

An exact quantification of spatial correlations in a liquid mixture is given by the Kirkwood-Buff integrals (KBI), which are defined by \cite{BenNaim2023}:
\begin{equation}
    G_{ij}=4\pi\int_o^\infty g_{ij}(r)r^2dr
\end{equation}
where $i$ and $j$ denote the species indices and $g_{ij}(r)$ is the corresponding radial distribution function between the centers of mass of species $i$ and $j$. The KBIs are related to the macroscopic thermodynamic properties in the $q\rightarrow0$ limit such as isothermal compressibility, activity coefficients, and partial molar volumes. Figure \ref{fig:kbi} shows the KBI versus $x_L$ at room temperature. All three sets of data are negative, with the cross KBI being the least negative and even slightly positive for $x_L=0.66$. This indicates that the unlike correlations dominate, thus exhibiting preferential hetero-association in the mixture. As the mixture becomes more saturated with L-menthol, the network between acid molecules becomes progressively disrupted, and for $x_L=0.66$ the packing of L-menthol around octanoic acid is maximized. This type of molecular organization is very different from that in aqueous mixtures of alcohols, such as ethanol and tert-butanol, where molecules show self-association leading to micro-heterogeneity \cite{Perera2006KirkwoodBuffAlcohol}. Negative KBI values are also observed in other H-bonded mixtures, such as the ethanol-methanol mixture, where local order appears in the form of chain-like structures \cite{Lovrincevic2019KirkwoodBuff}. Previous simulation studies \cite{Malik2021} show that there exists a low $q$-prepeak in the total structure factor $S(q)$ indicating a strong self-segregation of the polar groups, which is also supported by our results.
Figure SI-5. of the SI shows KBI values for temperatures $T=15\:^\circ C$ (left panel) and $T=35\:^\circ C$ (right panel). System snapshots for mixtures with $x_L=0.25, 0.50$ and $0.75$ shown in Figure SI-6. of the SI suggest that the two species mix almost perfectly in all concentrations.

\begin{figure}
    \centering
    \begin{subfigure}{0.45\textwidth}
        \centering
        \includegraphics[width=\linewidth]{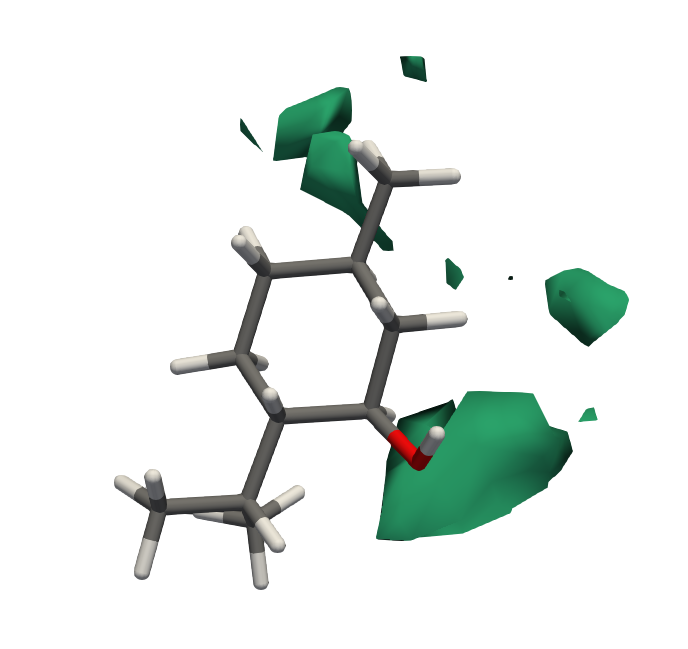}
        \caption{}
    \end{subfigure}
    \hfill
    \begin{subfigure}{0.45\textwidth}
        \centering
        \includegraphics[width=\linewidth]{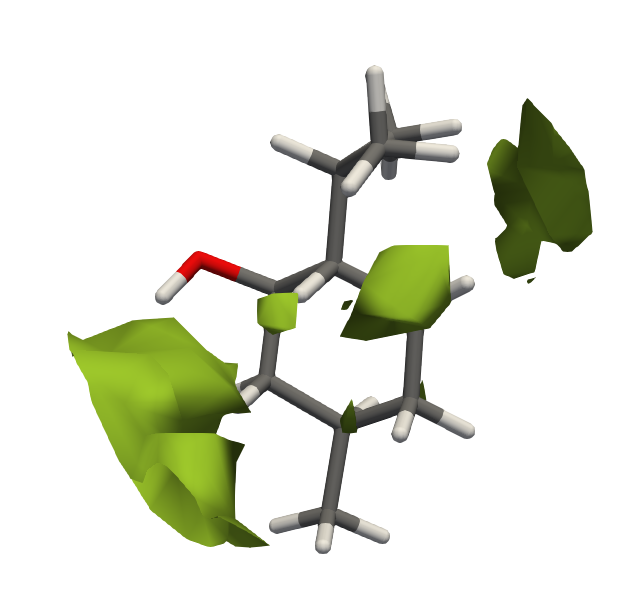}
        \caption{}
    \end{subfigure}

    \vspace{0.5cm}

    \begin{subfigure}{0.45\textwidth}
        \centering
        \includegraphics[width=\linewidth]{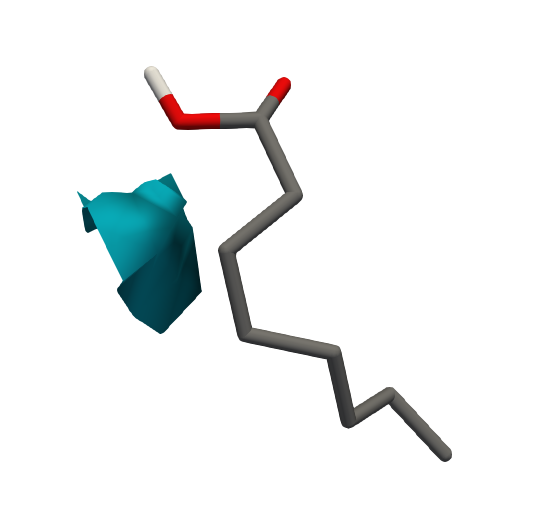}
        \caption{}
    \end{subfigure}
    \hfill
    \begin{subfigure}{0.45\textwidth}
        \centering
        \includegraphics[width=\linewidth]{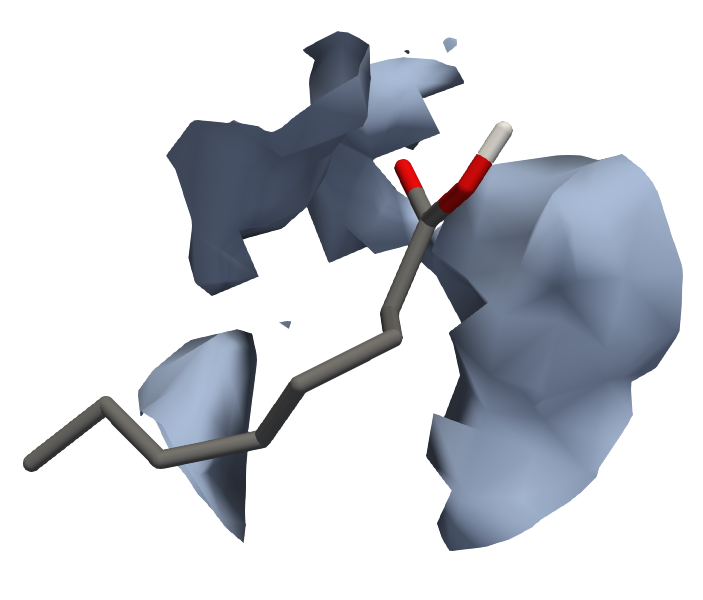}
        \caption{}
    \end{subfigure}

    \caption{Spatial distribution functions (SDF) around a single molecule at $T=25\:^\circ C$ for $x_L=0.50$. SDF of L-menthol molecules around L-menthol (a) and of octanoic acid around L-menthol (b). SDF of L-menthol molecules around one octanoic acid (c) and of acid molecules around one octanoic acid (d).}
    \label{fig:combined}
\end{figure}

We complement the KBI scenario and the $g(r)$s with the spatial distribution functions (SDF) shown in Figure \ref{fig:combined}, which reveal the actual spatial geometry of molecules surrounding a single molecule.  First, we analyze the SDF of L-menthol around a selected L-menthol molecule, as presented in panel Figure\ref{fig:combined}(a). We observe strong molecular localization around the -OH site as a result of H-bond formation with another L-menthol.  An isosurface near the H atom of the cycloalkane suggests the existence of the same H-bond. Figure \ref{fig:combined}(b) shows SDF of acid molecules around L-menthol. Again, the isosurface near the -OH group indicates H-bonding, as well as less prominent patches near the H atoms. Figure \ref{fig:combined}(c) shows the spatial density distribution of L-menthol near octanoic acid, where the surface near the O atom indicates the presence of the H-bond. Finally, in Figure \ref{fig:combined}(d) we observe that the isosurfaces of acid molecules follow the molecular shape close to the -OH group, as a result of both charge and steric effects. 

\subsubsection{Diffusion and hydrogen bond dynamics}
\begin{figure}
    \centering
    \includegraphics[width=0.9\linewidth]{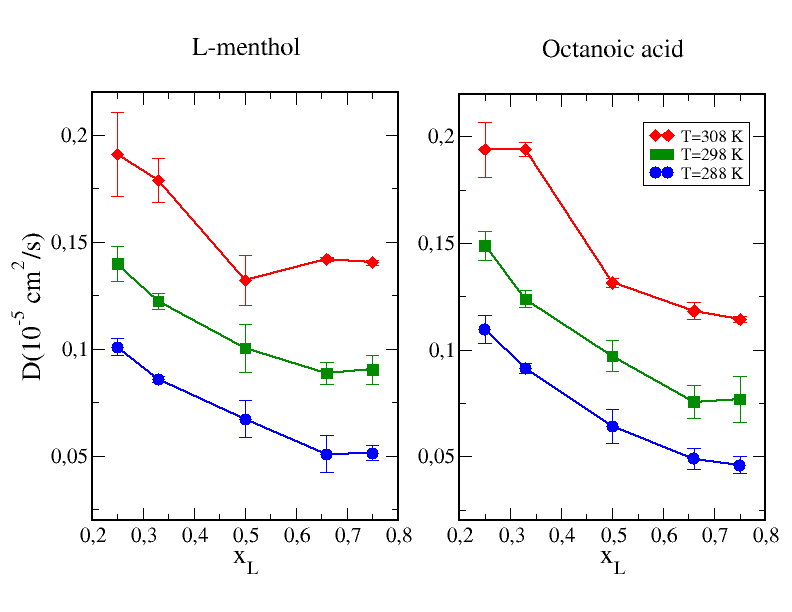}
    \caption{Self-diffusion coefficients $D\:(10^{-5}\:cm^2/s)$ vs. molar ratio $x_L$ for temperatures $T=15\:^\circ C$ (blue circles), $T=25\:^\circ C$ (dark green squares) and $T=35\:^\circ C$ (red diamonds). Left panel: L-menthol, right panel: octanoic acid.}
    \label{fig:diff}
\end{figure}

Self-diffusion coefficients $D$ for L-menthol (left panel) and octanoic acid (right panel) at all three temperatures are presented in Figure \ref{fig:diff}. The self-diffusion coefficients have been calculated by using the Einstein relation \cite{HansMac}:
\begin{equation}
    D=\frac{1}{6}\lim_{t\rightarrow \infty}\frac{d}{dt}\left<|r_i(t)-r_i(0)|^2 \right>
\end{equation}
where $i$ is the species index, $r_i(t)$ is the position of the molecular center-of-mass and the average $\left<\right>$ is taken over all molecules and all time origins.\\ 
Self-diffusion coefficients $D$ for both species decay with increasing $x_L$, which means that the diffusion of L-menthol is slower as this species becomes dominant in the system. This is consistent with the increasing viscosity in the system, which hinders translational motion of both molecules. Values of self-diffusion coefficients increase as the system becomes hotter, as molecules gain more kinetic energy which allows them to spread out more rapidly. When we compare the values of $D$ for L-menthol (left panel) and octanoic acid (right panel), we see that both diffuse almost equally fast, with L-menthol being slightly faster than octanoic acid in the alcohol-rich side of the mixture, probably due to steric effects. 

\begin{figure}
    \centering
    \includegraphics[width=0.9\linewidth]{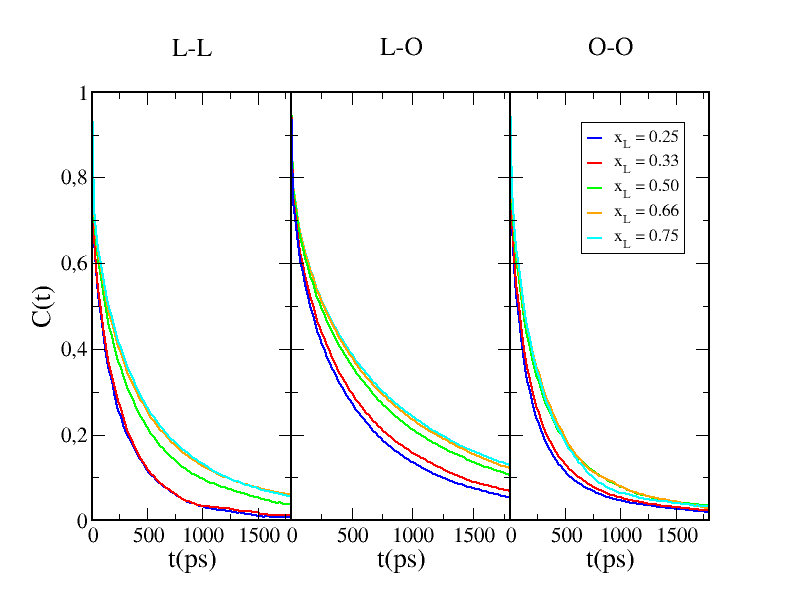}
    \caption{Hydrogen bond autocorrelation function $C(t)$ vs. $t(ps)$ at $T=25\:^\circ C$. Left panel: bond between two L-menthol molecules, middle panel: bond between L-menthol and octanoic acid, right panel: bond between two octanoic-acid molecules. Color convention: $x_L=0.75$ (cyan), $x_L=0.66$ (orange),$x_L=0.50$ (green),$x_L=0.33$ (red), and $x_L=0.25$ (blue). }
    \label{fig:Hbac}
\end{figure}

\begin{figure}
    \centering
    \includegraphics[width=0.9\linewidth]{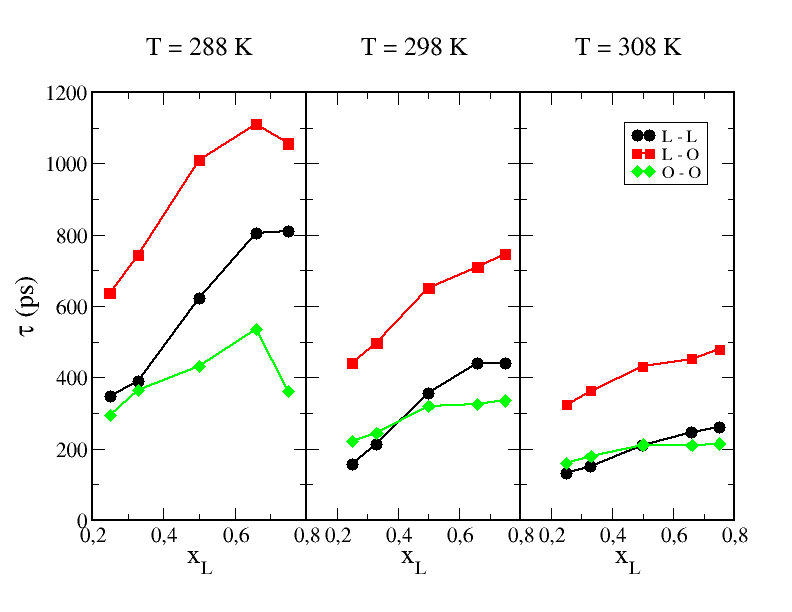}
    \caption{Hydrogen bond lifetimes $\tau_{HB}(ps)$ vs. $x_L$ Left panel: $T=15\:^\circ C$, middle panel:  $T=25\:^\circ C$, right panel:  $T=35\:^\circ C$. Color convention: L-menthol-L-menthol (black circles), L-menthol-octanoic acid (red squares) and octanoic acid-octanoic acid (green diamonds). }
    \label{fig:tauHB}
\end{figure}

Finally, we discuss the hydrogen bond dynamics. The hydrogen bond autocorrelation function $C_{HB}(t)$ is defined as \cite{Luzar1996, Luzar2000}:
\begin{equation}
    C_{HB}(t)=\frac{\left<h(0)h(t)\right>}{\left<h\right>}
\end{equation}
where $h(t)$ equals unity if the particular tagged pair of molecules is hydrogen bonded and zero otherwise. We use a three-exponential fit for the hydrogen bond autocorrelation function $C_{\mathrm{HB}}(t)$:
\begin{equation}
    C_{HB}(t)=\sum_{i=1}^3 A_ie^{-t/\tau_i}
\end{equation}
where $A_i$ is the amplitude of the $i$th decay process and $\tau_i$ is the corresponding time constant. The average lifetime of the hydrogen bond \cite{Balasubramanian2002, Rog2009} is then 
\begin{equation}
    \tau_{HB}=\sum_{i=1}^3A_i\tau_i
\end{equation}
The hydrogen bond autocorrelation function $C_{HB}(t)$ for all three types of bonds and for all mole fractions is shown in Figure \ref{fig:Hbac}. We focus on the decay of these functions, as they indicate the average lifetime of the hydrogen bond between two molecules. We keep in mind that the L-menthol molecule acts both as a single hydrogen bond donor and acceptor. The molecule of octanoic acid, on the other hand, can participate in the hydrogen bond as a double acceptor and a single donor. As seen in the right panel of Fig.\ref{fig:Hbac}, $C_{HB}(t)$ between two L-menthol molecules decay slower as there is more L-menthol in the system. The same is observed for the hydrogen bond between the two octanoic acid molecules (right panel of Fig\ref{fig:Hbac}), which suggests that L-menthol has a stabilizing effect on the overall hydrogen bond dynamics. The bond between L-menthol and the octanoic acid is the strongest, as indicated by the long decay of $C_{HB}(t)$ in the middle panel of Fig.\ref{fig:Hbac}.\\
The average hydrogen bond lifetimes $\tau_{HB}$ are shown in Figure \ref{fig:tauHB} for all three types of bonds and all the temperatures in the study. It is clear that the value of $\tau_{HB}$ between the L-menthol and the octanoic acid across all $x_L$ is the largest from the set for all $T$, as expected from the results in Fig.\ref{fig:Hbac}. The decrease in $T$ promotes the longevity of the hydrogen bond, as the values are the highest for $T=15\:^\circ C$ and then become smaller for $T=25\:^\circ C$ and $T=35\:^\circ C$. When comparing acid-acid and L-menthol-L-menthol bonds, there is a crossover between the values of $\tau_{HB}$ at $x_L\approx0.50$ for $T=25\:^\circ C$ (middle panel) and $T=35\:^\circ C$ (right panel). L-menthol molecules bond more strongly to the octanoic acid than to each other, so at small $x_L$ the bond between two L-menthol molecules is the weakest. Our findings are in line with previously reported MD results on menthol-based DES where the system stability relies on the intermolecular bonding between the menthol -OH group and the carboxylic acid -COOH \cite{Paul2020}. 

\section{Conclusion}

In this work, we performed a comprehensive experimental and molecular dynamics study of a hydrophobic eutectic solvent composed of L-menthol and octanoic acid over a range of compositions and temperatures. Five mixtures with molar ratios from 1:3 to 3:1 were investigated to elucidate the relationship between composition, thermophysical properties, and microscopic structure. 

Experimental measurements of density and viscosity were combined with molecular dynamics simulations using the OPLS force field, allowing a complementary analysis of macroscopic properties with molecular-level structure and dynamics. 
The experimental results show that the density decreases slightly with an increase in the L-menthol fraction, while the viscosity increases significantly as the system becomes richer in L-menthol.

Structural analysis based on radial distribution functions, 2D combined distribution functions and Kirkwood-Buff integrals reveals that the dominant structural motif in the liquid is the hydrogen bonding between the hydroxyl group of L-menthol and the carboxyl group of octanoic acid. These heteromolecular interactions are more prevalent and longer-lived than those between identical species, indicating that the stability of the mixture is largely governed by intermolecular L-menthol–octanoic acid hydrogen bonds.

Dynamic properties further support this structural picture. The self-diffusion coefficients decrease with increasing L-menthol content, reflecting the higher viscosity and reduced molecular mobility in L-menthol-rich mixtures. Analysis of hydrogen-bond autocorrelation functions shows that hydrogen bonds between L-menthol and octanoic acid exhibit the longest lifetimes among the three possible bonding types. This strong heteromolecular hydrogen-bond network plays a key role in stabilizing the liquid structure and contributes to the reduced melting point characteristic of deep eutectic solvents.

Overall, the combined experimental and simulation approach provides detailed information on the microscopic structure, transport properties, and hydrogen-bond dynamics of the L-menthol/octanoic acid hydrophobic eutectic solvent. The results demonstrate that the physicochemical properties of this system are strongly governed by intermolecular hydrogen bonding and molecular packing effects. These findings contribute to a better understanding of menthol-based ES systems and may support their further development in green extraction processes, pharmaceutical formulations, and environmentally friendly solvent technologies.

\section*{Acknowledgements}
This research was funded by the Ministry of Science, Education and Youth of the Republic of Croatia and the Slovenian Research and Innovation Agency (Croatian-Slovenian collaboration 2025./2026.) under the project name "Analysis of deep eutectic solvents as innovative and eco-friendly solvents in industrial application".
\section*{Conflict of Interest}
There is no conflict of interest to declare.
\bibliographystyle{unsrt}  
\bibliography{des}
\end{document}